 \documentclass[final,3p]{elsarticle}

 \usepackage{graphics}
 \usepackage{graphicx}
 \usepackage{epsfig}

\usepackage{amssymb}
 \usepackage{amsthm}
\usepackage{wasysym}

\makeatletter
\def\gsim{\compoundrel>\over\sim}
\def\lsim{\compoundrel<\over\sim}
\def\compoundrel#1\over#2{\mathpalette\compoundreL{{#1}\over{#2}}}
\def\compoundreL#1#2{\compoundREL#1#2}
\def\compoundREL#1#2\over#3{\mathrel
         {\vcenter{\hbox{$\m@th\buildrel{#1#2}\over{#1#3}$}}}}
\makeatother

\journal{Astroparticle Physics}

\begin{document}

\begin{frontmatter}



 \title{Search for neutrino emission from relic dark matter in the Sun with
  the Baikal NT200 detector}

%\title{Title}

%%\author[label1,label2]{<author name>}

%\author{author}

%\address{address}

\author[a]{A.D. Avrorin}
\author[a]{A.V. Avrorin}
\author[a]{V.M. Aynutdinov}
\author[g]{R. Bannasch}
\author[b]{I.A. Belolaptikov}
\author[c]{D.Yu. Bogorodsky}
\author[b]{V.B. Brudanin}
\author[c]{N.M. Budnev}
\author[a]{I.A. Danilchenko}
%S.\,Demidov$^{a}$\/\thanks{e-mail: demidov@ms2.inr.ac.ru},
\author[a]{S.V. Demidov}
\author[a]{G.V. Domogatsky}
\author[a]{A.A. Doroshenko}
\author[c]{A.N. Dyachok}
%Zh.-A.M.\,Dzhilkibaev$^{a}$\/\thanks{e-mail: djilkib@pcbai10.inr.ruhep.ru},
\author[a]{Zh.-A.M. Dzhilkibaev}
\author[e]{S.V. Fialkovsky}
\author[c]{A.R. Gafarov}
\author[a]{O.N. Gaponenko}
\author[a]{K.V. Golubkov}
\author[c]{T.I. Gress}
\author[b]{Z. Honz}
\author[g]{K.G. Kebkal}
\author[g]{O.G. Kebkal}
\author[b]{K.V. Konischev}
\author[c]{E.N. Konstantinov}
\author[c]{A.V. Korobchenko}
\author[a]{A.P. Koshechkin}
\author[a]{F.K. Koshel}
\author[d]{A.V. Kozhin}
\author[e]{V.F. Kulepov}
\author[a]{D.A. Kuleshov}
\author[a]{V.I. Ljashuk}
\author[e]{M.B. Milenin}
\author[c]{R.A. Mirgazov}
\author[d]{E.R. Osipova}
\author[a]{A.I. Panfilov}
\author[c]{L.V. Pan'kov}
\author[c]{A.A. Perevalov}
\author[b]{E.N. Pliskovsky}
\author[c]{V.A. Poleschuk}
\author[f]{M.I. Rozanov}
\author[c]{V.F. Rubtzov}
\author[c]{E.V. Rjabov}
\author[b]{B.A. Shaybonov}
\author[a]{A.A. Sheifler}
\author[d]{A.V. Shkurihin}
\author[b]{A.A. Smagina}
\author[a]{O.V. Suvorova\corref{cor1}}
\cortext[cor1]{Corresponding author}
\ead{suvorova@cpc.inr.ac.ru}
\author[c]{B.A. Tarashansky}
\author[g]{S.A. Yakovlev}
\author[c]{A.V. Zagorodnikov}
\author[a]{V.A. Zhukov}
\author[c]{V.L. Zurbanov}

%\address[a]{\scriptsize{Institute for Nuclear Research, 60th October Anniversary pr. 7A, Moscow 117312, Russia}}\vspace*{0.15cm}
\address[a]{Institute for Nuclear Research, 60th October Anniversary pr. 7A, Moscow 117312, Russia}
\address[b]{Joint Institute for Nuclear Research, Dubna 141980, Russia}
\address[c]{ Irkutsk State University, Irkutsk 664003, Russia}
\address[d]{ Skobeltsyn Institute of Nuclear Physics  MSU, Moscow 119991, Russia}
\address[e]{ Nizhni Novgorod State Technical University, Nizhni Novgorod 603950, Russia}
\address[f]{ St. Petersburg State Marine University, St. Petersburg 190008, Russia}
\address[g]{ EvoLogics GmbH, Berlin, Germany}

\begin{abstract}
We have analyzed a data set taken over 2.76 years live time 
with the Baikal neutrino telescope NT200. The goal of the analysis 
is to search for neutrinos
from dark matter annihilation in the center
of the Sun. Apart from the conventional annihilation channels $b\bar{b}$,
$W^+W^-$ and $\tau^+\tau^-$  we consider also the annihilation of
dark matter particles into monochromatic neutrinos. From the absence of any
excess of events from the direction of the Sun over the expected background, 
we derive 90\% upper limits on the fluxes
of muons and muon neutrinos from the Sun, as well as on the elastic cross sections 
of dark matter scattering on protons.
\end{abstract}


\end{frontmatter}
% \linenumbers

%%\section{}
%%\label{}
\section{Introduction}
\label{sec:introduction}
According to numerous astronomical and cosmological observations,
there exists a non-luminous form of matter which is responsible for
about 90\% of the mass within galaxies and clusters of galaxies.
Results from cosmic microwave background measurements by both
space- and ground-based experiments (e.g.,
\cite{WMAP2003},~\cite{WMAP2012},~\cite{Planck2013},~\cite{SouthPoleTelescope})
are remarkably consistent with the predictions of the standard
cosmological model in which the constituents 
of a spatially-flat and expanding Universe are dominated by cold dark
matter (CDM) and a cosmological constant ($\Lambda$) at later
times ($\Lambda$CDM model). Moreover, recent direct observations of
several merging clusters support a collisionless dark matter
scenario~\cite{Abel520}.
New collisionless dark matter (DM) particles can naturally appear in models 
beyond the Standard Model (SM). Supposed to be massive DM particles could be 
captured by astrophysical bodies (including the Sun) and accumulate 
inside them over cosmological times.
The most interesting range of DM particle masses for this scenario -- from
a few GeV to hundreds of TeV -- naturally arises in many classes of
theoretical models. Among the possible candidates for CDM of this kind of particles 
are the lightest superpartners in supersymmetric 
models~\cite{Jungman:1995df}, Kaluza-Klein particles in models with
extra dimensions~\cite{Hooper:2007qk} or 4th generation heavy
neutrinos~\cite{Khlopov}.
Through self-interactions, DM pairs
could annihilate into ordinary particles, and among the final products of
these annihilations could be high energy neutrinos~\cite{Griest:1986yu}.  Those are able
to reach the Earth and could be observed over the background of
atmospheric neutrinos. 
High energy neutrino telescopes (NTs~\cite{Katz:2012}) like the present and future
deep underwater detectors in Lake
Baikal~\cite{BaikalAstro:2011,GVD2013} can identify possible sources
of clumped dark matter if they observe a significant excess in the
number of neutrino events from the direction of the dark matter
accumulation. 

In this paper we analyse neutrino-induced upward going muons which
have been  measured with the Baikal neutrino telescope NT200
between April 1998 and February 2003. 
We study correlations of the arrival
directions with the Sun annual path.
Some preliminary results can be found in
Ref.~\cite{dmBaikal2009}. Here we perform a completely new
analysis of those data. We consider several possible annihilation
channels, assuming a 100\% branching ratio for each 
channel as extreme cases. 
The annihilation channels differ by the neutrino energy spectra 
and their flavour content produced from generated particles decays.
Soft spectra are dominated by energies much smaller than the mass of
the DM particles and emerge e.g. from $b\bar{b}$ decays, hard spectra
emerge from decays into $\tau^{+}\tau^{-}$ or $W^{+}W^{-}$ pairs.
To fully acknowledge flavour content of neutrino spectra
we also consider the annihilation of dark matter into monochromatic
neutrinos of all possibles flavours.
Another reason why we pay attention to monochromatic neutrinos is that 
they represent examples of the most energetic neutrino emissions in  
DM annihilations (we also refer reader to the analysis of
monochromatic neutrino signal in NTs performed in
Refs.~\cite{Farzan2012,Farzan2011}). We numerically optimized the
sizes of the half cones toward the Sun in dependence on to the assumed
mass of the DM particle. Finally, we compare our results with other experiments
looking for neutrino signal from DM annihilations in the Sun as well
as with the results of direct searches (for review see e.g.~\cite{Bergstrom2012}).

\section{Experiment and data selection}
\label{sec:experiment}
The neutrino telescope NT200 was completed in 1998. It is one of 
 several deep
underwater installations operating now in the southern basin of Lake
Baikal,  at a distance of 3.5 km off the shore and at a depth of \mbox{1.1
km}. Its goals are the study of high energy muons and neutrinos coming from the top and
bottom hemispheres, respectively. The NT200 configuration and its main functional
systems have been described in details
elsewhere~\cite{Baikal2007,Baikal2009}. The detector consists of 192
optical modules arranged pair-wise on 8 strings of 72\,m length:
seven peripheral strings and a central one. The distances between the
strings are about 20\,m. Each OM contains hybrid photodetector
QUASAR-370, a photo multiplier tube (PMT) with 37-cm diameter. To
suppress background from bioluminescence and dark noise, the two PMTs
of a pair are switched in coincidence within a time window of 15\,ns. 
Since 2005, the NT200 configuration has been upgraded by additional 3 strings 
each 100\,m away from the centre.
This upgraded detector, named NT200+, 
served as a prototype cell on the 10 Mton scale for a future Gigaton volume
detector~\cite{GVD2013} whose first cluster of 5 strings is operating since February 2014~\cite{GVD2014} . 

Relativistic particles crossing the effective
volume of a deep underwater telescope are detected via their
Cherenkov radiation. This radiation is recorded by optical modules (OMs) 
which are time-synchronized and
energy-calibrated by artificial light pulses.
At 1 km depth, the muon flux from cosmic ray interactions in the upper
hemisphere is about one million times higher than the flux of upward going muons
initiated by neutrino interactions in water and rock below the array. 
Our analysis is based on
data selection and event reconstruction described in details
previously, as in Ref.~\cite{BaikalAstro:2011,Belolap07}. We recall
them here briefly. The muon trigger requires $N_{hit}\geq n$ within 500 ns,
where hit refers to a pair of OMs coupled in a channel and \emph{n} is
set to 3 or 4. The first cut level selects events with
at least 6 hits on at least 3 strings ("6/3"), retaining about 40\% of
all triggered events.
Given the huge ratio of downward to upward moving muons,
the selection of a clean sample of true upward muons is
a major challenge and requires a highly efficient rejection of misreconstructed
downward moving muons. Therefore, a number of quality cuts is applied
during off-line data processing~\cite{BaikalAstro:2011,Belolap07}. 
Most fake events (downward muons mis-reconstructed as
upward muons) are due to muon bundles and populate directions close to the
horizon. Therefore only events with zenith angles
larger than $100^\circ$ were selected. The average rate of such events
was 0.037 Hz. Further optimization of the data selection criteria was
performed using simulated neutrino events~\cite{Belolap07} with a spectrum
following~\cite{Bartol}, as well as events from downward going
atmospheric muons generated with the help of 
CORSIKA~\cite{CORSIKA} and propagated with the MUM code~\cite{MUM}. 
To get the best possible
estimator in reconstruction of a muon trajectory, multiple start
guesses for the $\chi^2$ minimization are used: 
\begin{equation}
\chi^{2}=\frac{1}{(N_{hit}-5)}
\sum_{i=1}^{N_{hit}}\frac{(T_i(\theta,\phi,u_0,v_0,t_0)-t_i)^2}
    {\sigma_{ti}^2}.  
\end{equation}
Here, $t_i$ are the measured times and $T_i$ are the times expected for 
a given track hypothesis, $\sigma_{ti}$ are the timing errors. A set
of parameters defining a track hypothesis is given by $\theta$ and
$\phi$ -- the zenith and azimuth angle of the track, respectively,
$u_0$ and $v_0$ -- the two coordinates of the track point closest to
the centre of the detector, and $t_0$ - the time the muon passes this
point. The next step is the application of quality cuts to variables like the
number of hit channels, the probabilities of fired channels to have
been hit or not; the actual position of the track with respect to the
detector centre; the minimal residual between expected and detected times 
for hit channels; the smoothness of the channel response probabilities along
the reconstructed track and $\chi^2$/d.o.f.~\cite{Belolap07}. These
selection criteria were designed and optimized for 
the separation of atmospheric
neutrinos from the background of fake events from atmospheric muons. 
They provide a rejection factor for atmospheric
muons of about $10^{-7}$, resulting in a neutrino energy threshold
of about 10 GeV and a mean value $4.3^\circ$  for
the distribution of mismatch angles (compared to $14.1^\circ$ after the
initial ``6/3'' selection). The r.m.s. angle is $4.7^\circ$ and the median value
$2.5^\circ$. Note that in the further search for the optimal size
of the signal cones towards the Sun direction
we use the shape of the angular distribution itself.

\begin{figure}
\centerline{
\includegraphics[width=0.5\textwidth,height=0.5\textwidth]{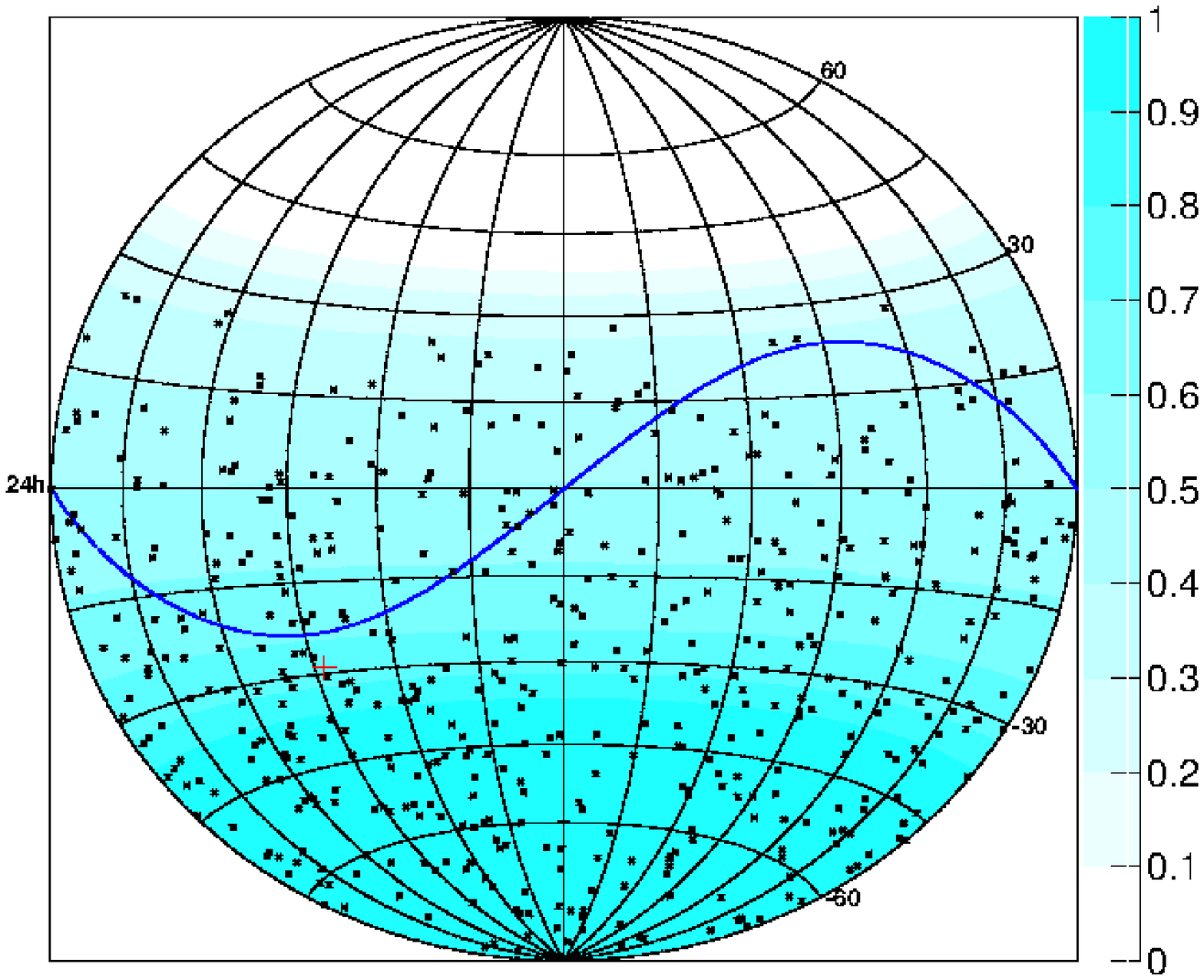}
}
\caption{\label{Fig1_skyNT200} 
Skymap with NT200 neutrino events in equatorial coordinates.
The blue curve marks the path of the Sun, the red cross --
the location of the Galactic Center.}
\end{figure}

\begin{figure}
\centerline{
\includegraphics[width=0.49\textwidth,height=0.49\textwidth]{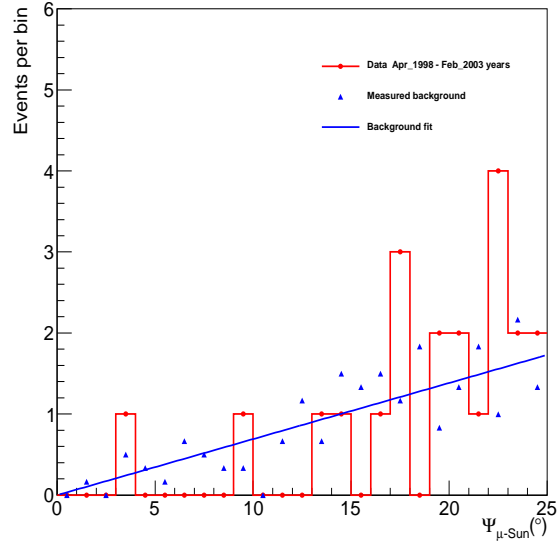}
}
%%\caption{\label{ris2_NT200Sun25} 
\caption{\label{Fig2_NT200Sun25} 
Data and backgound samples of mismatch anlges to the Sun.}
\end{figure}

The NT200 coordinates are $51.83^\circ$N and $104.33^\circ$E. 
The skyplot of the final sample of neutrino events 
in equatorial coordinates is shown in
Fig~\ref{Fig1_skyNT200}. The colour gradient marks the
sky visibility for NT200, the line follows the track of the Sun
averaged over a year. The total number of selected events surviving all 
cuts~\cite{dmBaikal2009},~\cite{Belolap07} is 510, for 1038 days of live time. 
We use a re-sampling (bootstrap)
method to get the background by random mixing of arrival directions and
times from the selected data sample. 
In Fig~\ref{Fig2_NT200Sun25} we show the numbers of observed events
(red histogram with points) 
and background (blue triangles) 
as a function of the angle between the Sun direction and the muon track.
The blue line is a fit to the background. 
We do not observe an excess of events from the direction of the Sun. 
Assuming Poisson statistics, the upper limits on the
number of signal events in excess of atmospheric neutrinos have been 
obtained at 90\% confidence level
(CL) applying the standard TRolke class in the ROOT analysis package~\cite{ROOT}. We leave 
the discussion of systematic uncertainties for Section~3. 
Next, based on this result, we derive other 
limits on parameters which can be easily compared with other experiments or 
with theoretical predictions.

\begin{figure}
\centerline{
\includegraphics[angle=-90,width=0.6\textwidth]{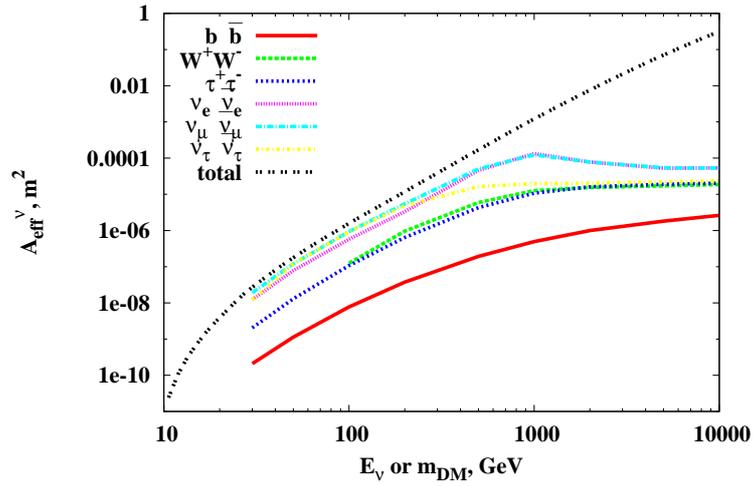}
}
\caption{\label{area} 
Effective area as a function of neutrino energy $E_{\nu}$
("total") and as a function of the mass $m_{DM}$ of the DM particle
for different annihilation
channels.}  
\end{figure}

For the search for signal events from the decay of annihilating DM
particles in the Sun we optimize the sizes of cones toward the Sun
in a way which gives the best signal-to-background ratio, taking into
account for each decay mode the spectrum for signal neutrinos coming at night.
For the calculation of the neutrino effective area $A_{eff, \nu}$,
we apply the NT200 response function (efficiency) for crossing muons
generated in charged current neutrino interactions. 
We have used the same MC mass production of neutrinos (antineutrinos) as in 
Ref.~\cite{Belolap07} and reconstructed muons
for twelve NT200 configurations operated 
in 1998-2003. The following expression describes these calculations:

\begin{equation}
\label{eqn : Seffnu}
A^{i}_{eff, \nu}(E_{th}, E_{\nu}) =  V^{i}_{MC} \times N_A \times \rho
\times \sigma^{CC}(E_{\nu}) \times  \epsilon^{i}_{\nu}(E_{th},
E_{\nu}),  
\end{equation}
where $i=1, ...,12$ and $\epsilon^{i}_{\nu}$ is the efficiency of muon
reconstruction i.e. the ratio of two-dimensional angular-energy
distributions of reconstructed events to simulated neutrinos. The MC
weights of each event were included and the Sun trajectory also. 
The mean value of the generated volumes $V^{i}_{MC}$ from 
all twelve configurations is $V_{MC}=4.406\times{10}^{14}$cm$^3$.
The product of
$N_A \cdot \rho \cdot \sigma^{CC}(E_{\nu})$ in (\ref{eqn : Seffnu}) is the
inverse value of the length of neutrino-nucleon interactions in the Earth.
The value $N_A$ is the Avogadro number,
$\rho$ the medium density (rock or water), ${\sigma^{CC}}$ 
the neutrino-nucleon cross
section in charged current (CC) interactions. The shadowing effect due
to the exponential attenuation of the neutrino flux in the Earth is negligible
or weak for energies less than 10 TeV and we can omit it in formula
(\ref{eqn : Seffnu}). The result for the neutrino effective area 
in dependence on the neutrino energy is shown in
Fig.~\ref{area} as a black line and labeled "total". 
The convolution of the effective area $A^{\nu}_{eff}$ with the 
given neutrino flux ${\Phi^{\nu}}$ 
gives the expected number of neutrinos either from a signal source or from background.
We optimize the signal to background ratio as presented below.

\section{Signal simulation and results}
In this Section we describe the results of the numerical simulation of the
neutrino signal from dark matter annihilations in the Sun. For the
calculations we use a numerical procedure which has been described in
Ref.~\cite{Boliev:2013ai}. Here we only briefly sketch the main
ingredients of these simulations.  

In any particular model the neutrino signal from dark matter
annihilations in the Sun depends on combinations of different
annihilation channels. For a model-independent approach we suppose
that dark matter annihilates over particular channels with 100\%
branching ratio and give separate limits for each case. 
As representative annihilation channels we have chosen
the conventional $b\bar{b}$, $\tau^{+}\tau^{-}$, $W^{+}W^{-}$ decays
and in addition the direct annihilation into neutrinos $\nu_e\bar{\nu}_e$,
$\nu_{\mu}\bar{\nu}_{\mu}$ and $\nu_{\tau}\bar{\nu}_{\tau}$. The
latter annihilation channels yield the most energetic neutrinos.
The neutrino spectra at the production point
have been calculated using the WimpSIM package~\cite{wimpsim,Blennow:2007tw}
which uses PYTHIA for this purpose. Subsequently we propagate
neutrinos from the Sun to the Earth, taking into account neutrino
oscillations in vacuum and in the media of the Sun and the Earth, and NC
and CC neutrino interactions of neutrinos with nucleons. We use the solar
model presented in Ref.~\cite{Bahcall:2004pz}. As neutrino
oscillation parameters we use $\Delta m_{21}^2 = 7.62\cdot
10^{-5}~{\rm eV}^2$, $\Delta m_{31}^{2} = 2.55\cdot 10^{-3}~{\rm
  eV}^2$, $\delta_{CP} = 0$, $\sin^2{\theta_{12}} = 0.32$,
$\sin^2{\theta_{23}} = 0.49$, $\sin^2{\theta_{13}} = 0.026$ which are
suggested by current experimental results~\cite{Tortola:2012te}.  Also
we take into account $\nu_{\tau}$ regeneration. The results of these
simulations are energy spectra $\frac{dN_{\nu_{\mu}}}{dE_{\nu_{\mu}}}$
for muon neutrinos and antineutrinos at the telescope location for
different annihilation channels and values of the mass of dark matter
particles. Using the neutrino energy spectra we simulate CC neutrino
interactions in  the media surrounding the telescope and 
propagate the muons
in rock and water. We refer the reader to
Ref.~\cite{Boliev:2013ai} for more details. We note in passing that in
Ref.~\cite{Baratella:2013fya} a more thorough calculation of initial and
final neutrino spectra has been performed which includes electroweak
corrections to the spectra. We do not take them into account because
we are going to present the comparison with the limits obtained by
other experiments which did not take them into account either.

\begin{figure}
\centerline{
\includegraphics[angle=-90,width=0.9\textwidth]{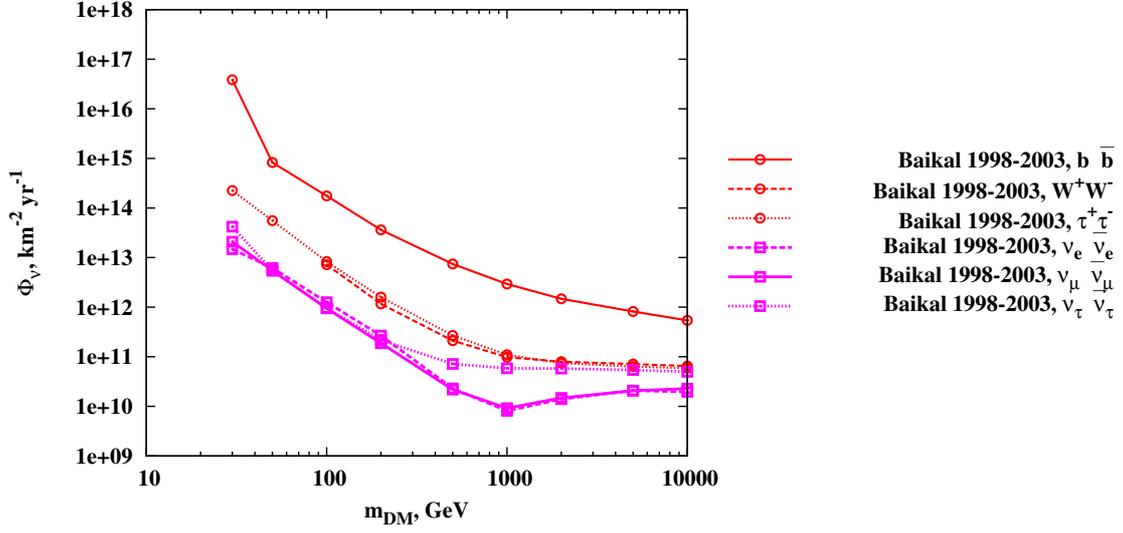}
}
\caption{\label{nu_flux} 
90\% CL upper limits on the muon neutrino flux for NT200.}
\end{figure}

\begin{figure}
\centerline{
\includegraphics[angle=-90,width=0.9\textwidth]{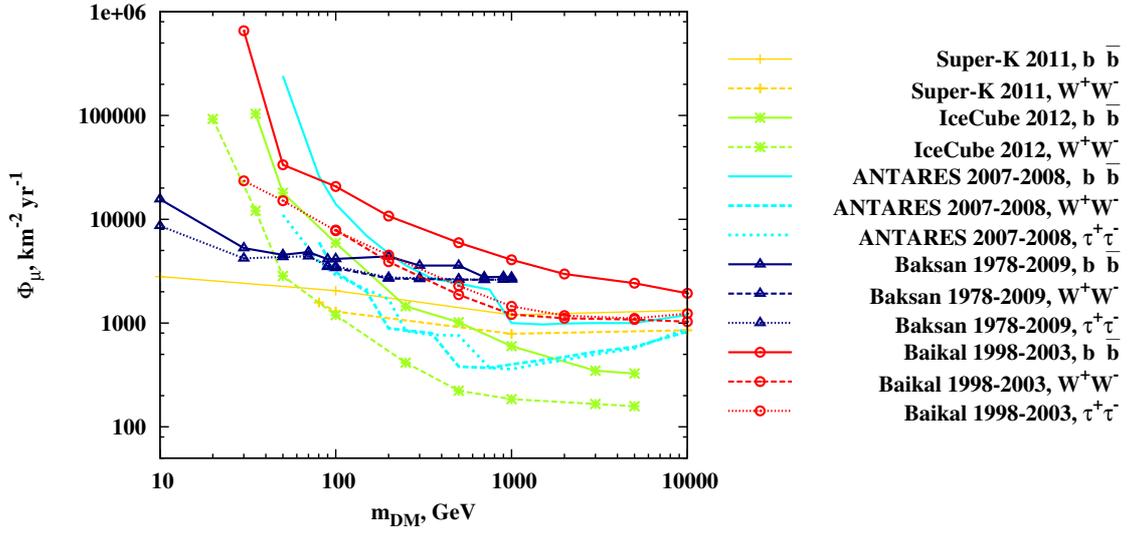}
}
\caption{\label{mu_flux} 
90\% CL upper limits on the muon flux for NT200 in comparison with the
results from other neutrino telescopes.}
\end{figure}

The Sun has an apparent size of about $0.5^\circ$ and can be
considered as a local point source. However, the expected muon signal
has a much larger angular spread because of the kinematics
of the initial CC neutrino interaction and
because of the intrinsic angular resolution $\Psi$ of the 
telescope. This rises the question about how to define the 
size of the cone half-angle around the
direction towards the Sun in which the events will be counted. We
optimize the values of this angle to
obtain the tightest expected upper limits on the neutrino flux.
For the optimization of the search cone sizes we follow the MRF
approach~\cite{Hill:2002nv}. For a given mass of the dark matter particle 
and for each annihilation channel we construct the expected limit
on the neutrino flux $\Phi_{\nu + \bar{\nu}}^{exp}$ as a function of the
cone half-angle $\psi$ as follows  
\begin{equation}
\label{mrf}
\Phi_{\nu + \bar{\nu}}^{exp}(\psi) =
\frac{\bar{N}_{S}^{90}(\psi)}{A_{eff}(\psi)\times T}.
\end{equation}
Here $\bar{N}_{S}^{90}$ is the 90\% C.L. upper limit on the number of
neutrino events inside the given cone averaged over the number of
signal events 
with a Poisson distribution, $T$ is the livetime. The effective area
$A_{eff}(\psi)$ is defined for a given angular window $\psi$ and the
muon neutrino energy spectrum $\frac{dN_{\nu_{\mu}}}{dE_{\nu_{\mu}}}$
at the telescope location as 
\begin{equation}
A_{eff}(\psi) = \frac{\sum_{\nu_{\mu},\bar{\nu}_{\mu}} \int_{E_{th}}^{m_{DM}} dE_{\nu_{\mu}}
A(E_{\nu_{\mu}},E_{th})  P(E_{\nu_{\mu}},\psi) 
  \frac{dN_{\nu_{\mu}}(E_{\nu_{\mu}})}{dE_{\nu_{\mu}}}
}{\sum_{\nu_{\mu},\bar{\nu}_{\mu}} \int_{E_{th}}^{m_{DM}}   dE_{\nu_{\mu}}
  P(E_{\nu_{\mu}},\psi) 
  \frac{dN_{\nu_{\mu}}(E_{\nu_{\mu}})}{dE_{\nu_{\mu}}}}.
\end{equation}
Here $A(E_{\nu_{\mu}},E_{th})$ is defined according to~(\ref{eqn : Seffnu}) 
and the sum over muon neutrino and antineutrino is 
implied. The probability $P(E_{\nu},\psi)$ for neutrinos with energy
$E_{\nu}$ to produce a muon which falls within the cone of
half-angular size $\psi$ from the initial neutrino direction has been
obtained as described above, taking into account the angular
resolution of the telescope obtained from MC simulation of the
detector response~\cite{Belolap07}. By minimizing the  
function~(\ref{mrf}) we find the optimal values of the half-cone
angle. The corresponding effective areas for different annihilation
channels are shown in Fig.~\ref{area}. As expected, $\nu\bar{\nu}$
annihilation channels lead to the largest effective areas and the
soft $b\bar{b}$ annihilation channels to the smallest ones.
The upper limit on the neutrino flux is obtained by Eq.~(\ref{mrf})
replacing $\bar{N}_{S}^{90}$ with values of the actual upper limit
$N_{S}^{90}$ discussed in the previous section. The upper limits on
the muon neutrino flux for all chosen annihilation channels, 
recalculated for a neutrino energy threshold of 1~GeV, 
are shown in Fig.~\ref{nu_flux} and listed
in Table~\ref{table:results}, for DM masses ranging from 30~GeV to 10~TeV.  
One can see that the direct neutrino channels result in the tightest
limits. The limits for the $b\bar{b}$, $\tau^{+}\tau^{-}$ and $W^{+}W^{-}$ 
channels are shown in in Fig.~\ref{mu_flux} as limits on the muon flux
and compared to results from other experiments. 

The upper limits on the muon neutrino flux can be related to the upper
limits on other physical quantities related to dark matter physics. 
First of all, the muon neutrino flux directly depends on the DM
annihilation rate $\Gamma_{A}$ in the Sun as follows
\begin{equation}
\Phi_{\nu_{\mu}} = \frac{\Gamma_{A}}{4\pi
  R^2}\sum_{\nu_{\mu},\bar{\nu}_{\mu}} \int_{E_{th}}^{m_{DM}} 
dE_{\nu_{\mu}} \frac{dN_{\nu_{\mu}}}{dE_{\nu_{\mu}}},
\end{equation}
where where $R$ is the distance to the Sun, and $E_{th}=10$~GeV
is the threshold neutrino energy of NT200. 
The upper limits on the annihilation rate $\Gamma_{A}$ for
several values of dark matter particle masses and different annihilation
channels is also presented in Table~\ref{table:results}.

After the long-time evolution of the solar system the processes of capture
and annihilation of dark matter in the Sun can reach an equilibrium. In
this case the capture rate should be twice the annihilation rate
and the upper limit on $\Gamma_{A}$ can be recalculated to limits
on the spin-dependent (SD, $\sigma^{UppLim}_{SD}$) and spin-independent
(SI, $\sigma^{UppLim}_{SI}$) parts of the cross section of elastic
scattering of dark matter on proton~\cite{Edsjo:09,Demidov:2010rq} as
follows 
\begin{eqnarray}
\label{recalc1}
\sigma^{UppLim, SD}_{\chi, p}(m_{\chi})= \lambda^{SD}\left(
m_{\chi}\right) \cdot \Gamma^{UppLim}_A(m_{\chi})\\ 
\label{recalc2}
\sigma^{UppLim, SI}_{\chi, p}(m_{\chi})= \lambda^{SI} \left( m_{\chi}\right) \cdot
\Gamma^{UppLim}_A(m_{\chi}).
\end{eqnarray}
For completeness we present formulas for the the coefficients
$\lambda^{SD}$ and $\lambda^{SI}$ in appendix~A, as described in
Refs.~\cite{Boliev:2013ai}.  Their values are obtained assuming $\rho_{DM} =
0.3$~GeV/cm$^{3}$ for the local dark matter density and $v_d =
270$~km/s for the root-mean-square of the dark matter velocity dispersion.  

Now let us discuss the systematic uncertainties of the obtained
results. The experimental uncertainty is estimated to be 30\% which
includes uncertainties in 
the optical properties of water and of the sensitivity of the OMs
(see details for example in Refs.~\cite{BaikalAstro:2011},~\cite{BaikalMonop:2008}). 
Among the main theoretical uncertainties we 
mention those related to neutrino properties (interactions and
oscillations). We estimate the uncertainties due to present errors in 
oscillation parameters as about 8\% for leptonic annihilation channels
and 5\% for $W^{+}W^{-}$ and $b\bar{b}$. 
The errors in
neutrino-nucleon cross sections 
results in uncertainties from 8\% to 4\% for dark matter masses 
ranging from 30~GeV to 10~TeV. 
The estimation has been done by varying neutrino oscillation
parameters~\cite{Tortola:2012te} or neutrino-nucleon cross
section~\cite{CooperSarkar:2011pa} within their known ranges of uncertainty. 
Other systematic uncertainties include nuclear formfactors, 
the solar chemical composition,
the influence of planets on dark matter capture by the solar system as
well as the dark matter velocity
distribution~\cite{Edsjo:09,Rott:2011fh,Rott:2014jc}. Let us mention that some 
recent studies indicate a larger value for the local dark matter
density around 0.4~GeV/cm$^3$~\cite{Catena:2009mf} which would make the
upper limits even stronger.

\begin{figure}
\centerline{
\includegraphics[angle=-90,width=0.9\textwidth]{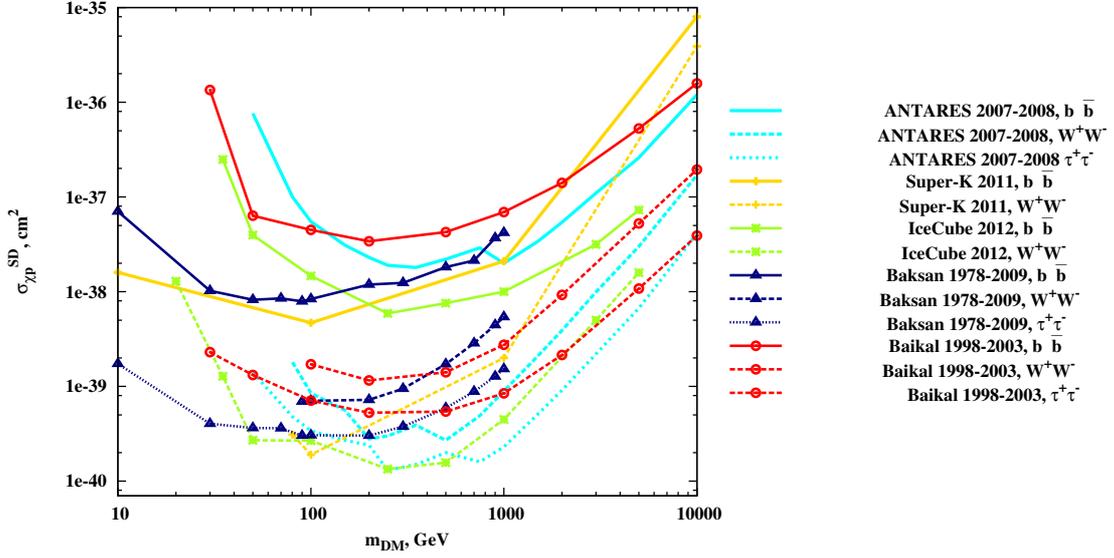}
}
\caption{\label{SD_indirect} 
Comparison of Baikal 90\% CL limits on the spin-dependent elastic cross section
of DM particles on protons with the results of other indirect searches
from ANTARES, Baksan, IceCube, SuperKamiokande.}
\end{figure}

\begin{figure}
\centerline{
\includegraphics[angle=-90,width=0.9\textwidth]{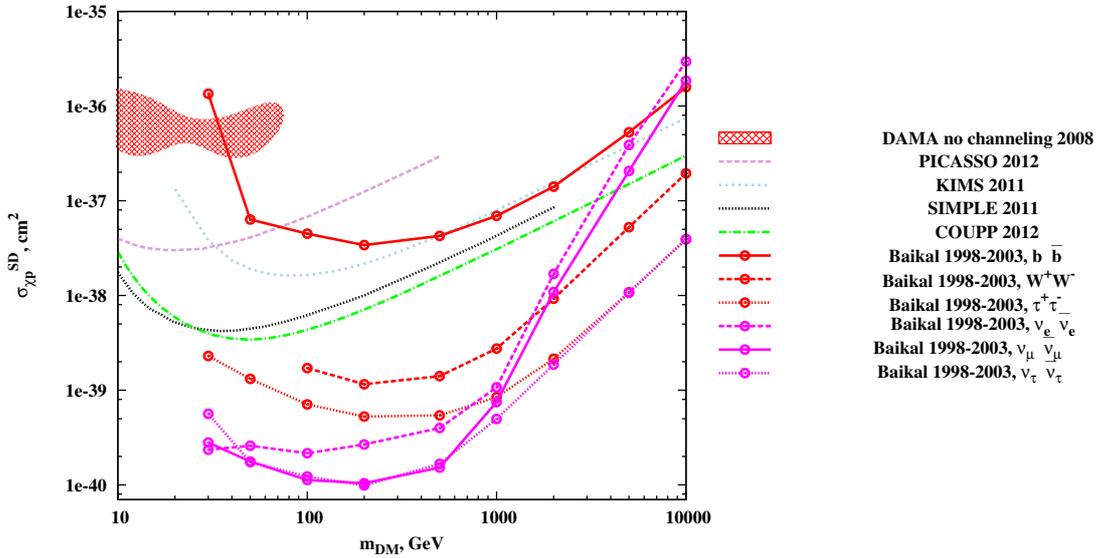}
}
\caption{\label{SD_direct} 
Comparison of Baikal 90\% CL limits on the spin-dependent elastic cross section
of DM particles on protons with the results of direct searches: DAMA,
PICASSO, KIMS, SIMPLE, COUPP.}
\end{figure}

The incorporation of all systematic uncertainties for signal and
background into the final upper limits have been performed by the standard
way~\cite{Beringer:1900zz} and the using TRolke class in ROOT. 
A full description of the method can be found in~\cite{trolke},
while for the extraction of the confidence intervals a likelihood function is used. 
Systematic uncertainties result in a degradation of the upper
limits by 7-12\%.  
The final results for half-cone angles
$\gamma$, upper 90\% limits on the number of signal  events
$N_{S}^{90}$,  the muon flux $\Phi_{\mu}$, the dark matter
annihilation rate in the Sun  $\Gamma_{A}$, the dark matter -proton
spin-dependent $\sigma^{SD}_{\chi p}$ and spin-independent
$\sigma^{SI}_{\chi p}$  scattering cross sections and neutrino fluxes
$\Phi_{\nu}$ are shown in Table~\ref{table:results}.
\begin{table}%TABLE{
\footnotesize
\begin{center}
\begin{tabular}{|ccccccccc|}
\hline\hline
$m_{\rm DM},${\rm GeV}$ $ & channel & $\gamma$, deg &
$N_s^{90}$ & $\Phi_{\mu}$,${\rm km}^{-2}/{\rm yr}$ &
$\Gamma_{A}$, ${\rm s}^{-1}$ & $\sigma^{SI}_{\chi p}$, pb & 
$\sigma^{SD}_{\chi p}$, pb  & $\Phi_{\nu_{\mu}}$,${\rm km}^{-2}/{\rm yr}$
\\\hline
30.0 & $b\bar{b}$ & 10.6 & 1.2 & $6.6\cdot 10^{5}$ & $1.4\cdot
10^{27}$ & $8.2\cdot 10^{-3}$ & $1.3$  & $3.8\cdot 10^{16}$\\
     & $\tau^{+}\tau^{-}$ & 9.2 & 0.57 & $2.3\cdot 10^4$ & $2.3\cdot
10^{23}$ & $1.4\cdot 10^{-5}$ & $2.3\cdot 10^{-3}$ & $2.3\cdot 10^{14}$  \\
     & $\nu_{e}\bar{\nu}_{e}$ & 9.3 & 0.50 & $9.2\cdot 10^3$ & $2.4\cdot
10^{23}$ & $1.4\cdot 10^{-6}$ & $2.4\cdot 10^{-4}$ & $1.5\cdot 10^{13}$ \\
     & $\nu_{\mu}\bar{\nu}_{\mu}$ & 8.5 & 1.1 & $1.2\cdot 10^4$ & $2.9\cdot
10^{23}$ & $1.7\cdot 10^{-6}$ & $2.8\cdot 10^{-4}$ & $2.1\cdot 10^{13}$  \\
     & $\nu_{\tau}\bar{\nu}_{\tau}$ & 8.1 & 1.3 & $2.3\cdot 10^4$ & $5.7\cdot
10^{23}$ & $3.4\cdot 10^{-6}$ & $5.6\cdot 10^{-4}$ & $4.2\cdot 10^{13}$  \\
\hline
50.0 & $b\bar{b}$ & 9.4 & 0.43 & $3.3\cdot 10^4$ & $2.8\cdot
10^{25}$ & $2.5\cdot 10^{-4}$ & $6.3\cdot 10^{-2}$ & $8.3\cdot 10^{14}$ \\
     & $\tau^{+}\tau^{-}$ & 8.2 & 1.3 & $1.5\cdot 10^4$ & $5.7\cdot
10^{23}$ & $5.2\cdot 10^{-6}$ & $1.3\cdot 10^{-3}$ & $5.6\cdot 10^{13}$ \\
     & $\nu_{e}\bar{\nu}_{e}$ & 8.1 & 1.3 & $9.9\cdot 10^3$ & $1.1\cdot
10^{23}$ & $1.0\cdot 10^{-6}$ & $2.6\cdot 10^{-4}$ & $6.2\cdot 10^{12}$  \\
     & $\nu_{\mu}\bar{\nu}_{\mu}$ & 7.5 & 1.7 & $8.1\cdot 10^3$ & $7.7\cdot
10^{22}$ & $7.0\cdot 10^{-7}$ & $1.8\cdot 10^{-4}$ & $5.5\cdot 10^{12}$  \\
     & $\nu_{\tau}\bar{\nu}_{\tau}$ & 7.5 & 1.7 & $8.2\cdot 10^3$ & $7.8\cdot
10^{22}$ & $6.9\cdot 10^{-7}$ & $1.7\cdot 10^{-4}$ & $5.4\cdot 10^{12}$  \\
\hline
100.0 & $b\bar{b}$ & 8.2 & 1.3 & $2.1\cdot 10^4$ & $5.5\cdot
10^{24}$ & $1.0\cdot 10^{-4}$ & $4.5\cdot 10^{-2}$ & $1.7\cdot 10^{14}$ \\
     & $\tau^{+}\tau^{-}$ & 7.2 & 1.9 & $7.8\cdot 10^3$ & $8.7\cdot
10^{22}$ & $1.6\cdot 10^{-6}$ & $7.1\cdot 10^{-4}$ & $8.3\cdot 10^{12}$ \\
     & $W^{+}W^{-}$ & 7.2 & 1.9 & $7.9\cdot 10^3$ & $2.1\cdot
10^{23}$ & $4.0\cdot 10^{-6}$ & $1.7\cdot 10^{-3}$ & $7.2\cdot 10^{12}$ \\
     & $\nu_{e}\bar{\nu}_{e}$ & 7.1 & 2.0 & $7.7\cdot 10^3$ & $2.7\cdot
10^{22}$ & $5.0\cdot 10^{-7}$ & $2.2\cdot 10^{-4}$ & $1.3\cdot 10^{12}$ \\
     & $\nu_{\mu}\bar{\nu}_{\mu}$ & 6.6 & 2.3 & $4.7\cdot 10^3$ & $1.4\cdot
10^{22}$ & $2.6\cdot 10^{-7}$ & $1.1\cdot 10^{-4}$ & $9.4\cdot 10^{11}$  \\
     & $\nu_{\tau}\bar{\nu}_{\tau}$ & 6.6 & 2.3 & $4.5\cdot 10^3$ & $1.2\cdot
10^{22}$ & $2.8\cdot 10^{-7}$ & $1.2\cdot 10^{-4}$ & $9.7\cdot 10^{11}$  \\
\hline
200.0 & $b\bar{b}$ & 7.4 & 1.8 & $1.1\cdot 10^4$ & $1.1\cdot
10^{24}$ & $5.0\cdot 10^{-5}$ & $3.4\cdot 10^{-2}$ & $3.6\cdot 10^{13}$ \\
     & $\tau^{+}\tau^{-}$ & 6.4 & 2.4 & $4.5\cdot 10^3$ & $1.7\cdot
10^{22}$ & $7.7\cdot 10^{-7}$ & $5.3\cdot 10^{-4}$ & $1.6\cdot 10^{12}$ \\
     & $W^{+}W^{-}$ & 6.3 & 2.4 & $3.9\cdot 10^3$ & $3.8\cdot
10^{22}$ & $1.3\cdot 10^{-6}$ & $1.4\cdot 10^{-3}$ & $2.1\cdot 10^{11}$ \\
     & $\nu_{e}\bar{\nu}_{e}$ & 6.3 & 2.4 & $4.5\cdot 10^3$ & $8.8\cdot
10^{21}$ & $3.9\cdot 10^{-7}$ & $2.8\cdot 10^{-4}$ & $2.7\cdot 10^{11}$  \\
     & $\nu_{\mu}\bar{\nu}_{\mu}$ & 6.0 & 2.6 & $2.5\cdot 10^3$ & $3.4\cdot
10^{21}$ & $1.5\cdot 10^{-7}$ & $1.0\cdot 10^{-4}$ & $1.9\cdot 10^{11}$  \\
     & $\nu_{\tau}\bar{\nu}_{\tau}$ & 6.0 & 2.6 & $2.6\cdot 10^3$ & $3.2\cdot
10^{21}$ & $1.5\cdot 10^{-7}$ & $9.9\cdot 10^{-5}$ & $2.2\cdot 10^{11}$  \\
\hline
500.0 & $b\bar{b}$ & 6.5 & 2.3 & $5.9\cdot 10^3$ & $2.3\cdot
10^{23}$ & $4.1\cdot 10^{-5}$ & $4.2\cdot 10^{-2}$ & $7.4\cdot 10^{12}$ \\
     & $\tau^{+}\tau^{-}$ & 5.7 & 2.7 & $2.3\cdot 10^3$ & $2.9\cdot
10^{21}$ & $5.2\cdot 10^{-7}$ & $5.4\cdot 10^{-4}$ & $2.7\cdot 10^{11}$ \\
     & $W^{+}W^{-}$ & 5.6 & 2.8 & $1.2\cdot 10^3$ & $7.6\cdot
10^{21}$ & $1.3\cdot 10^{-6}$ & $1.4\cdot 10^{-3}$ & $2.1\cdot 10^{11}$ \\
     & $\nu_{e}\bar{\nu}_{e}$ & 5.5 & 2.8 & $1.1\cdot 10^3$ & $2.2\cdot
10^{21}$ & $3.8\cdot 10^{-7}$ & $4.0\cdot 10^{-4}$ & $2.3\cdot 10^{10}$  \\
     & $\nu_{\mu}\bar{\nu}_{\mu}$ & 5.5 & 2.8 & $1.0\cdot 10^3$ & $8.2\cdot
10^{20}$ & $1.5\cdot 10^{-7}$ & $1.5\cdot 10^{-4}$ & $2.2\cdot 10^{10}$ \\
     & $\nu_{\tau}\bar{\nu}_{\tau}$ & 5.5 & 2.8 & $1.4\cdot 10^3$ & $9.0\cdot
10^{20}$ & $1.6\cdot 10^{-7}$ & $1.7\cdot 10^{-4}$ & $7.2\cdot 10^{10}$ \\
\hline
1000.0 & $b\bar{b}$ & 6.1 & 2.5 & $4.1\cdot 10^3$ & $9.5\cdot
10^{22}$ & $5.5\cdot 10^{-5}$ & $6.9\cdot 10^{-2}$ & $2.9\cdot 10^{12}$ \\
     & $\tau^{+}\tau^{-}$ & 5.5 & 2.8 & $1.4\cdot 10^3$ & $1.2\cdot
10^{21}$ & $6.8\cdot 10^{-7}$ & $8.4\cdot 10^{-4}$ & $1.1\cdot 10^{11}$ \\
     & $W^{+}W^{-}$ & 5.5 & 2.8 & $1.2\cdot 10^3$ & $3.8\cdot
10^{21}$ & $2.2\cdot 10^{-6}$ & $2.7\cdot 10^{-3}$ & $9.8\cdot 10^{10}$ \\
     & $\nu_{e}\bar{\nu}_{e}$ & 5.3 & 2.9 & $6.7\cdot 10^2$ & $1.5\cdot
10^{21}$ & $8.6\cdot 10^{-7}$ & $1.1\cdot 10^{-3}$ & $8.1\cdot 10^{9}$  \\
     & $\nu_{\mu}\bar{\nu}_{\mu}$ & 5.3 & 2.9 & $5.2\cdot 10^2$ & $1.0\cdot
10^{21}$ & $6.0\cdot 10^{-7}$ & $7.5\cdot 10^{-4}$ & $9.4\cdot 10^{9}$  \\
     & $\nu_{\tau}\bar{\nu}_{\tau}$ & 5.5 & 2.8 & $1.1\cdot 10^3$ & $6.8\cdot
10^{20}$ & $4.0\cdot 10^{-7}$ & $5.0\cdot 10^{-4}$ & $5.8\cdot 10^{10}$  \\
\hline
2000.0 & $b\bar{b}$ & 5.8 & 2.7 & $3.0\cdot 10^3$ & $7.4\cdot
10^{20}$ & $1.0\cdot 10^{-4}$ & $0.14$ & $1.5\cdot 10^{12}$ \\
     & $\tau^{+}\tau^{-}$ & 5.4 & 2.9 & $1.2\cdot 10^3$ & $1.2\cdot
10^{21}$ & $1.6\cdot 10^{-6}$ & $2.1\cdot 10^{-3}$ & $7.5\cdot 10^{10}$ \\
     & $W^{+}W^{-}$ & 5.4 & 2.9 & $1.1\cdot 10^3$ & $3.2\cdot
10^{21}$ & $6.8\cdot 10^{-6}$ & $9.2\cdot 10^{-3}$ & $8.0\cdot 10^{10}$ \\
     & $\nu_{e}\bar{\nu}_{e}$ & 5.3 & 2.9 & $7.5\cdot 10^2$ & $5.8\cdot
10^{21}$ & $1.2\cdot 10^{-5}$ & $1.7\cdot 10^{-2}$ & $1.4\cdot 10^{10}$  \\
     & $\nu_{\mu}\bar{\nu}_{\mu}$ & 5.3 & 2.9 & $6.9\cdot 10^2$ & $3.7\cdot
10^{21}$ & $7.9\cdot 10^{-6}$ & $1.1\cdot 10^{-2}$ & $1.5\cdot 10^{10}$  \\
     & $\nu_{\tau}\bar{\nu}_{\tau}$ & 5.4 & 2.9 & $1.1\cdot 10^3$ & $6.4\cdot
10^{20}$ & $1.3\cdot 10^{-6}$ & $1.9\cdot 10^{-3}$ & $5.8\cdot 10^{10}$  \\
\hline
5000.0 & $b\bar{b}$ & 5.6 & 2.8 & $2.4\cdot 10^3$ & $2.9\cdot
10^{22}$ & $3.7\cdot 10^{-4}$ & $0.53$ & $8.2\cdot 10^{11}$\\
     & $\tau^{+}\tau^{-}$ & 5.4 & 2.9 & $1.1\cdot 10^3$ & $5.9\cdot
10^{20}$ & $7.5\cdot 10^{-6}$ & $1.1\cdot 10^{-2}$ & $6.3\cdot 10^{10}$ \\
     & $W^{+}W^{-}$ & 5.4 & 2.9 & $1.1\cdot 10^3$ & $2.9\cdot
10^{21}$ & $3.6\cdot 10^{-5}$ & $5.3\cdot 10^{-2}$& $7.1\cdot 10^{10}$  \\
     & $\nu_{e}\bar{\nu}_{e}$ & 5.3 & 2.9 & $9.5\cdot 10^2$ & $2.1\cdot
10^{22}$ & $2.7\cdot 10^{-4}$ & $3.8\cdot 10^{-1}$ & $2.1\cdot 10^{10}$  \\
     & $\nu_{\mu}\bar{\nu}_{\mu}$ & 5.6 & 2.8 & $7.1\cdot 10^2$ & $1.1\cdot
10^{22}$ & $1.4\cdot 10^{-4}$ & $2.1\cdot 10^{-1}$ & $2.1\cdot 10^{10}$  \\
     & $\nu_{\tau}\bar{\nu}_{\tau}$ & 5.3 & 2.9 & $1.1\cdot 10^3$ & $5.9\cdot
10^{20}$ & $7.4\cdot 10^{-6}$ & $1.1\cdot 10^{-2}$ & $5.4\cdot 10^{10}$  \\
\hline
10000.0 & $b\bar{b}$ & 5.7 & 2.7 & $1.9\cdot 10^3$ & $2.2\cdot
10^{22}$ & $1.1\cdot 10^{-3}$ & $1.58$ & $5.4\cdot 10^{11}$ \\
     & $\tau^{+}\tau^{-}$ & 5.4 & 2.9 & $1.2\cdot 10^3$ & $5.4\cdot
10^{20}$ & $2.7\cdot 10^{-5}$ & $3.9\cdot 10^{-2}$ & $5.9\cdot 10^{10}$ \\
     & $W^{+}W^{-}$ & 5.4 & 2.9 & $1.0\cdot 10^3$ & $2.7\cdot
10^{21}$ & $1.3\cdot 10^{-4}$ & $0.19$ & $6.5\cdot 10^{10}$ \\
     & $\nu_{e}\bar{\nu}_{e}$ & 5.3 & 2.8 & $8.8\cdot 10^2$ & $4.1\cdot
10^{22}$ & $2.0\cdot 10^{-3}$ & 2.96 & $2.0\cdot 10^{10}$  \\
     & $\nu_{\mu}\bar{\nu}_{\mu}$ & 5.3 & 2.9 & $7.7\cdot 10^2$ & $2.5\cdot
10^{22}$ & $1.3\cdot 10^{-3}$ & $1.8$ & $2.3\cdot 10^{10}$  \\
     & $\nu_{\tau}\bar{\nu}_{\tau}$ & 5.4 & 2.9 & $1.0\cdot 10^3$ & $5.4\cdot
10^{20}$ & $2.7\cdot 10^{-5}$ & $3.9\cdot 10^{-2}$ & $5.0\cdot 10^{10}$  \\
\hline\hline
\end{tabular}
\end{center}
\caption{\label{table:results}
 Half-cone angles $\gamma$, 90\% upper limits  $N_{S}^{90}$ on the number of signal
 events,  the muon flux $\Phi_{\mu}$, the dark matter
 annihilation rate in the Sun $\Gamma_{A}$, the dark matter-proton
 spin-dependent  $\sigma^{SD}_{\chi p}$ and spin-independent
 $\sigma^{SI}_{\chi p}$ scattering cross sections and neutrino
 fluxes $\Phi_{\nu}$.} 
\end{table}
%}
The fluxes have been recalculated to an energy threshold of 1~GeV.
In Fig.~\ref{SD_indirect} we present the Baikal upper limits on SD
cross section for the annihilation channels $b\bar{b}$, $W^+W^-$ and
$\tau^+\tau^-$  (red lines) in comparison with the results 
of indirect searches by the ANTARES~\cite{Adrian-Martinez:2013ayv},
Baksan~\cite{Boliev:2013ai}, IceCube~\cite{:2012ef}(where ``hard'' means either
$\tau^+\tau^-$ or $W^+W^-$ channel correspondingly to DM masses which could be
lighter or heavier than W-boson) and
Super-Kamionande~\cite{Tanaka:2011uf} collaborations. Obviously,
each experiment has better sensitivity for harder neutrino spectra from 
$\tau^+\tau^-$ annihilation channel. 
In Fig.~\ref{SD_direct} we compare the Baikal upper limits for all six
chosen annihilation channels including annihilations into monochromatic
neutrinos with the results of direct searches by
PICASSO~\cite{Archambault:2012pm}, KIMS~\cite{Kim:2012rza},
SIMPLE~\cite{Felizardo:2011uw}, DAMA~\cite{DAMA:08,Savage:2008er} and
COUPP~\cite{Behnke:2010xt}. 
Obviously, the neutrino telescope bounds are complementary to direct
detection results. The upper limits on SD and SI cross sections
obtained with NT200 results refer to 100\% branching ratio of the
corresponding annihilation channels. In general model one expects that 
these upper bounds lie somewhere between those annihilation channels for 
the softest and hardest neutrino spectra if DM particles annihilate over them
with a considerable fraction. Otherwise the limits are weakened by
corresponding branching ratio. This should be beared in mind when
comparing them with the results of direct searches. 
Let us comment at this point about the
behaviour of the upper limits for direct neutrino annihilation
channels (pink lines). Clearly, these limits are considerably stronger
than those for $b\bar{b}$, $W^+W^-$ and $\tau^+\tau^-$ channels (red
lines) for $m_{DM}\lsim  
700$~GeV. This is related to the fact that the neutrino spectrum at
production in the Sun for $\nu\bar{\nu}$ channels is more energetic and
a harder neutrino energy spectrum is expected at the telescope location. At
the same time for larger masses $m_{DM}\gsim 700$~GeV a suppression in
the upper limits arises because of increasing CC and NC
neutrino-nucleon cross sections which result in considerable absorption
of $\nu_e$ and $\nu_{\mu}$ and their antineutrinos in the Sun. 
However, because of the effect of $\nu_{\tau}$ regeneration, the
corresponding neutrinos reappear after CC interactions with somewhat
lower energy. Therefore the limits for $\tau^{+}\tau^{-}$ and
$\nu_{\tau}\bar{\nu}_{\tau}$ annihilations channels are quite similar
for very large DM masses. These observations indicate that 
indirect searches with neutrino telescopes are more sensitive to
leptophilic dark matter models than to others. 

\section{Conclusions}
\label{sec:conclusions}
To summarize, we have studied the NT200 response to neutrinos
from dark matter annihilations in the
Sun. We have derived upper limits on the muon
and neutrino fluxes, the annihilation rate and on the SD/SI cross sections 
of DM scattering on protons assuming different annihilation channels. We have
found that for DM masses below 500~GeV the best sensitivity is obtained
for the annihilation in neutrino-antineutrino pairs.
For heavier dark matter particles the most stringent limits are
obtained for leptonic $\tau^+\tau^-$ and $\nu_{\tau}$
annihilation channels. 

\section*{Acknowledgments}

{\footnotesize
%\begin{acknowledgments}
%The results were presented at the International workshop
%EPNT2013\footnote{https://indico.in2p3.fr/event/epnt13}. We thanks the organizers for a fruitful meeting. 
We are grateful to C. Spiering for valuable comments.
The research was supported in part by RFBR grant 13-02-12221, by RFBR grant 14-02-00972, by RFBR grant 13-02-01127 (S.D.) and 
by grant of the President of the Russian Federation NS-2835.2014.2
(S.D.).}   
%\end{acknowledgments}


%% References
%%   \citet{key}         ==>   Author [#]


\bibliographystyle{model6-num-names}
\bibliography{<your-bib-database>}

%% not want to use model6-num-names.bst.


%%
%%  // All plots
%%
%%++ Figure:1

%%++ Figure:2

%%++ Figure:4

%%++ Figure:4

%%++ Figure:4

%%++ Figure:3

%%++ Figure:4

%%++ Figure:4
\begin{figure}
\centerline{
\includegraphics[angle=-90,width=0.9\textwidth]{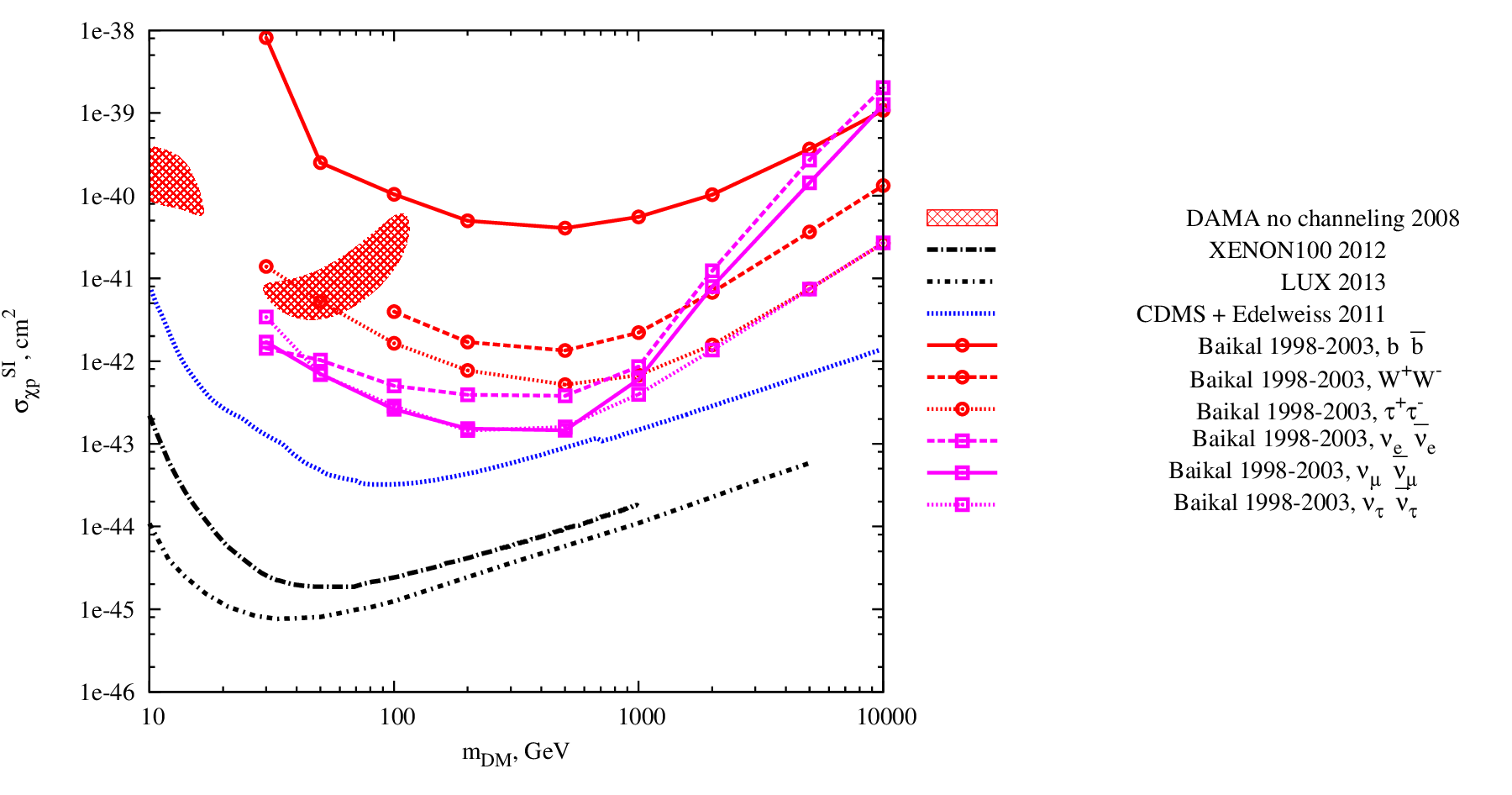}
}
\caption{\label{SI_direct} 
Comparison of Baikal 90\% CL limits on the spin-independent elastic cross section
of DM particles on protons with the results of direct searches: DAMA,
CDMS/Edelweiss, XENON, LUX.}
\end{figure}

\appendix

\section{Calculation of $\lambda^{SD}$ and $\lambda^{SI}$}
In this Appendix we present the procedure used to calculate the
coefficients $\lambda^{SD}$ and $\lambda^{SI}$ entering
Eqs.~(\ref{recalc1}) and~(\ref{recalc2}). They can be obtained from known
expressions~\cite{Gould:1992} (see also~\cite{Edsjo:09}) for dark
matter capture rate $C$ and the equilibrium condition. The capture
rate can be represented in the form of the following integral
over the Sun volume
\begin{equation}
\label{capture1}
C = \int_{0}^{R_{\astrosun}}4\pi r^2dr \sum_{i}\frac{dC_i}{dV},
\end{equation}
where $R_{\astrosun}$ is the radius of the Sun and the sum goes over
different types of nuclei in the Sun. The integrand in
(\ref{capture1}) is obtained by averaging of the capture probability
over velocities $u$ of dark matter particles as  follows
\begin{equation}
\label{capture}
\frac{dC_i}{dV} = \int_{0}^{u_{max}}du
\frac{f(u)}{u}\left(w\Omega_{v,i}(w)\right). 
\end{equation}
Here 
\begin{equation}
u_{max} = 2v_{esc}\frac{\sqrt{\mu}}{\mu - 1},\;\;\;
\mu = \frac{m_{\chi}}{m_{i}}, \;\;\;
w = \sqrt{u^2 + v_{esc}^2},
\end{equation}
$v_{esc}$ is escape velocity, $m_{i}$ is the mass of $i$-th type
of nuclei and the velocity distribution has the form 
\begin{equation}
\frac{f(u)}{u} = \frac{3}{2\pi}\frac{\rho_{\chi}}{m_{\chi}v_{d}v_{\astrosun}}
\left({\rm exp}\left(-\frac{3(u-v_{\astrosun})^2}{2v_d^2}\right)
-{\rm exp}\left(-\frac{3(u+v_{\astrosun})^2}{2v_d^2}\right)\right),
\end{equation}
where $v_{d} = 270$~km/s is velocity dispersion, $v_{Sun} = 220$~km/s
is the velocity of the Sun relative to the dark matter halo,
$\rho_{\chi}= 0.3$~GeV/cm$^3$ is the local dark matter density. The
capture probability per time unit and for $i$-th type of element
$\Omega_{v,i}$ which enter Eq.~(\ref{capture}) 
can be expressed as follows 
\begin{equation}
w\Omega_{v,i}(w) =
\sigma_{{\chi},i}n_i %\frac{\rho_{i}}{m_{i}}
\frac{(\mu+1)^2}{2\mu} \frac{E_{i}^0}{m_{\chi}}\left[{\rm 
    exp}\left(-\frac{m_{\chi}u^2}{2E_{i}^0}\right) - {\rm
    exp}\left(-\frac{2\mu}{(\mu+1)^2} \frac{m_{\chi}w^2}{E_{i}^0}\right)\right],    
\end{equation}
where $\sigma_{\chi,i}$ is the elastic cross section of the dark matter
on the $i$-th type of nuclei, $n_i$ is number density of the $i$-th
  element and the exponential suppression with a coherence energy  
\begin{equation}
E_{i}^0 = \frac{3 {\hbar}^2}{2m_{\chi}R^2_{i}},\;\;\;\;\; 
R_{i} = \left[0.91\left(\frac{m_i}{\rm
    GeV}\right)^{1/3} + 0.3\right]\cdot 10^{-15}~{\rm m}
\end{equation}
comes from a ``form factor'' in the cross section of dark
matter particle on nuclei. For the SD part of the cross section
$\sigma^{SD}$ we take into account only scattering on hydrogen. As for
the SI part of the cross section on the $i$-th type of nuclei it can be
calculated as  
\begin{equation}
\sigma^{SI}_{\chi, i} = \sigma^{SI}A^2_{i}
\frac{(m_{\chi}m_{i})^2}{(m_{\chi} + m_{i})^2}\frac{(m_{\chi} +
  m_{p})^2}{(m_{\chi}m_{p})^2}
\end{equation}
where $\sigma^{SI}$  is the SI cross section of dark matter on proton,
$A_{i}$ is the atomic number and $m_p$ is the proton mass. The expression
for the capture rate can be divided into spin-dependent and
spin-independent parts   
\begin{equation}
C = C^{SD} + C^{SI}
\end{equation}
according to which type of cross section contributes to it.
 Finally, the coefficients entering~(\ref{recalc1})
  and~(\ref{recalc2}) can be expressed as follows
\begin{equation}
\lambda^{SD}(m_{\chi}) = \frac{\sigma^{SD}}{2 C^{SD}}, \;\;\;\;
\lambda^{SI}(m_{\chi}) = \frac{\sigma^{SI}}{2 C^{SI}},
\end{equation}
where the factor of $2$ comes from the condition of equilibrium between
capture and annihilation processes which implies
%$C=\frac{1}{2}\Gamma_{A}$. 
2$C=\Gamma_{A}$ . 
To obtain $\lambda^{SD}(m_{\chi})$ and
$\lambda^{SI}(m_{\chi})$ we perform a numerical integration in
Eqs.~(\ref{capture1}) and~(\ref{capture}) using the solar model BS2005-OP
of Ref.~\cite{Bahcall:2004pz}. For elements heavier than oxygen the
relative abundances were taken from Ref.~\cite{Grevesse:1998bj}. 

\end{document}